\documentclass{svjour2}                    
\smartqed  
\usepackage{amsmath,amssymb}
\usepackage{graphicx}
\def\ii{{i}}

\def\ee{\mathrm e}

\def\tr{\mathrm {tr}}

\def\SL{\mathrm{SL}}

\def\Sp{\mathrm{Sp}}
\def\Mp{\mathrm{Mp}}

\def\Hom{\mathrm{Hom}}

\def\cS{\mathcal {S}}
\def\cW{\mathcal {W}}

\def\CC{{\mathbb C}}
\def\ZZ{{\mathbb Z}}
\def\RR{{\mathbb R}}

\def\QQ{{\mathbb Q}}

\def\fg{{\mathfrak{g}}}
\def\ft{{\mathfrak{t}}}

\def\ind{{\mathrm {ind}}}
\def\intf{{\mathrm {intf}}}

\def\Aut{{\mathrm {Aut}}}
\def\sign{{\mathrm {sgn}}}

\def\cA{{\mathcal{A}}}

\def\hx{{\hat{x}}}
\def\hw{{\hat{w}}}
\def\hu{{\hat{u}}}

\def\hA{{\hat{A}}}
\def\hB{{\hat{B}}}
\def\hC{{\hat{C}}}
\def\hD{{\hat{D}}}
\def\hphi{{\hat{\phi}}}
\def\hkappa{{\hat{\kappa}}}

\def\book#1{\rm{#1}, }
\def\paper#1{\textit{#1}, }
\def\jour#1{\rm{#1}, }
\def\yr#1{({\rm{#1})}}
\def\vol#1{\textbf{#1}}
\def\pages#1{\rm{#1}}

\def\publaddr#1{\rm{#1}, }
\def\publ#1{\rm{#1}, }
\def\by#1{{\rm{#1}, }}

\begin{document}

\title{Gauss Optics and Gauss Sum on an Optical Phenomena
}
\author{Shigeki MATSUTANI}



\institute{S. Matsutani \at
              8-21-1 Higashi-Linkan, Sagamihara 228-0811 \\
              \email{rxb01142@nifty.com}           
}

\date{Received: date / Accepted: date}

\maketitle

\begin{abstract}
In the previous article (Found Phys. Lett. {\bf{16}} 325-341),
we showed that a reciprocity of the Gauss sums is connected with the
wave and particle complementary.
In this article, we revise the previous investigation
by considering a relation between the Gauss optics and
the Gauss sum based upon the recent studies of
the Weil representation for a finite group.
\end{abstract}

\section{Introduction}
\label{intro}

In the previous article \cite{MO}, we investigated 
a relation between the Gauss reciprocity and wave-particle complementary
on an optical system, the fractional 
Talbot system \cite{WW1} following the excellent
work of Berry and Klein \cite{BK}.

More precisely, in \cite{MO}, we considered the wavy and 
particle-like treatments
of the system, or treatments based upon the Helmholtz equation
and the Fresnel integral.  Both treatments express the same phenomena
and the final results must agree; the agreement between
them means a kind of the wave-particle complementary.
As the fractional Talbot phenomena has discrete nature,
the distribution at the screen expressed in terms of
the Gauss sum parameterized by coprime integers $(p, q)$ \cite{BK};
the Gauss sum is a number theoretical function
which is well-known in number theory (see \S \ref{subsec:PNR})
and plays the central roles in both
quadratic number theory and cyclotomic field theory \cite{IR}.
Corresponding to the agreement between 
 the wavy and particle-like treatments,
there appears a reciprocity between the Gauss sums parameterized
by both $(p, q)$ and $(q, p)$ \cite{MO}, which is known as
the Gauss reciprocity \cite[Chap.8]{He}.

In \cite{MO} even though we dealt with the optical system,
using the similarity between para-axial optics and non-relativistic
quantum mechanics \cite[p.75-84]{GS},
we have concluded that in the system,
the canonical commutation relation in the quantum mechanics,
\begin{equation}
	q p  - p q= \ii \hbar
\label{eq:CCR}
\end{equation}
for position operator $q$ and momentum operator $p$,
is similar to a relation in the primitive number
theory that for coprime numbers $p$ and $q$ 
there exist integers 
 $\left\{\dfrac{1}{q}\right\}_{p}$ and $\left\{-\dfrac{1}{p}\right\}_{q}$ 
such that
\begin{equation}
	p \left\{\frac{1}{p}\right\}_q
	-q \left\{-\frac{1}{q}\right\}_p = 1.
\label{eq:PNR}
\end{equation}
The existence of these integers is primitively proved, {\it{e.g.}},
\cite[Theorem III.1]{Wei2} \cite[p.4, Lemma 4]{IR}
(see \S \ref{subsec:PNR}).
Most of all theorems in number theory are based upon the relation
(\ref{eq:PNR}) and it is regarded as a fundamental relation in 
number theory. It is very similar to the fact that (\ref{eq:CCR}) is the
fundamental relation in the quantum mechanics.
Both (\ref{eq:CCR}) and (\ref{eq:PNR})
play the central roles in the wave-particle complementary 
and the Gauss reciprocity respectively.
In \cite{MO} a question arises why they appear
and play the similar roles in the optical system.
The purpose of this article is to answer this question.

On the other hand, in \cite{Wei1},
Weil studied the symplectic group and
the Gauss sum based upon development of the quantum mechanics,
and found the Heisenberg group, its Schr\"odinger representation 
and a unitary representation of the
metaplectic group known as the Weil representation.
The Weil representation is essential to the foundation of 
quantum mechanics and linear optics.
Following \cite{Wei1}, 
Guillemin and Sternberg \cite{GS} and
Raszillier and Schempp \cite{RS} gave physical interpretations
of the Heisenberg group, the Schr\"odinger representation 
and the Weil representation in quantum mechanics and optics. 
Due to the discreteness and finiteness properties of the fractional
Talbot system, the Heisenberg group related to the system
becomes a finite group as we will show in \S \ref{sec:revised}.
Recently the relations between the Weil representation and the Gauss sum 
over a finite ring are studied well \cite{Bl,CMS,Sc,Sz,Tu}.
In this article, we answer the question in \cite{MO}
following these studies.

The fractional Talbot phenomena also has some influence on
modern optics, quantum problems and materials physics, {\it e.g.},
 optical fibers \cite{RTCM}, 
 quantum information \cite{BCSG}, the cyclotomic quantum clock problem
\cite{RP}, and composite metamaterials
such as multilayer positive and negative optical index media \cite{FHO}.
It is crucial to understand the algebraic essential of
the fractional Talbot phenomena in terms of the Heisenberg group
and the Weil representation.  Then it enables us to answer a question
how the discrete nature and the property of the finiteness
influence the optical system.

Though it is historical irony, K. F. Gauss studied well the Gauss sum 
and wrote it in \cite{G1} 1818 and summarized the Gauss optics in
\cite{G2} 1840, in which he described the critical relation between 
$\Sp(2, \RR) = \SL(2, \RR)$ and optical system, whereas H. F. Talbot
discovered the original Talbot phenomena around 1836 \cite{Ta1}, though
the fractional Talbot phenomena was discovered by 
J. T. Winthrop and C. R. Worthington in \cite{WW1}.
Talbot also studied elliptic integrals
as a mathematician \cite[p.413]{C}
and wrote several articles in the creation of the 
theory of elliptic functions \cite{WW2}.
For example, he wrote about the {\lq\lq}Abelian integral{\rq\rq} in a letter
of Sept. 8, 1844 to J. F. W. Herschel \cite{Ta2}.
It is well-known that Gauss also studied elliptic integrals and
elliptic functions alone \cite{C}.
The periodicity and algebraic structure in the elliptic integrals
were the theme in the elliptic function theory. 
I will show that in the Talbot phenomena,
the periodicity and algebraic structure also plays important role.
In fact the Gauss sum and the optical system are formulated by
the elliptic theta functions as we will show in (\ref{eq:phi2I,II,theta}).
Further 
 recent studies \cite{De,J,Tu} show that the generalized theta functions and 
Gauss sum are connected in other physical problems {\it{e.g.}},
 Chern-Simons-Witten theory.
 Thus to consider  relations among Gauss sum,
Gauss optics, elliptic theta function and fractional Talbot phenomena is
also interesting from the viewpoints of recent developments and 
science history.

Here we mention the contents of this article; Section 2 is devoted
to the mathematical preliminary of  the fundamental relation
(\ref{eq:PNR}), the Gauss sum and the Gauss reciprocity. 
Section 3 is a review of the
Gauss Optics and $\SL(2, \RR)$ based upon \cite{GS}. 
In section 4, following \cite{KK}, we give the fractional Talbot phenomena
in the framework of the Gauss optics and show the relation between
the Gauss sum and Gauss optics.
Section 5 is devoted to a review of the Heisenberg group and Weil
representation following \cite{GS,RS} and the recent movements 
 \cite{Bl,CMS,Sc,Sz,Tu}.
In section 6, as a revised investigation of the fractional Talbot phenomena
in \cite{MO}, we discuss these properties,
especially algebraic properties of the the fractional Talbot phenomena.
There we find a key fact to an answer of the question in \cite{MO}.
In section 7, we will give an answer to the question using
the key fact and further comments.

Here $\RR$, $\QQ$, and $\ZZ$ denote sets of real numbers,
the fractional numbers, and integers respectively.

\section{Mathematical Preliminary}

This section is for the mathematical preliminary of  the fundamental relation
(\ref{eq:PNR}), the Gauss sum and the Gauss reciprocity. 

\subsection{On the fundamental relation (\ref{eq:PNR})}
\label{subsec:PNR}

In number theory, it is well known that
there exist integers 
 $\left\{\dfrac{1}{q}\right\}_{p}$ and $\left\{-\dfrac{1}{p}\right\}_{q}$ 
as in (\ref{eq:p_q_modq}) such that
\begin{equation}
	p \left\{\frac{1}{p}\right\}_q
	-q \left\{-\frac{1}{q}\right\}_p = 1.
\label{eq:PNR2}
\end{equation}
This can be easily and elementally proved; for example see
\cite[Theorem III.1]{Wei2} and \cite[p.4, Lemma 4]{IR}.
This relation is a theoretical base of primitive  number theory
\cite{IR,Wei2}.

From (\ref{eq:PNR2}), there uniquely exists
a positive integer $\left[\dfrac{1}{p}\right]_{q}$ 
smaller than $q$ satisfying
\begin{gather}
p \left[\dfrac{1}{p}\right]_{q}\equiv 1
\text{  mod }q, 
\label{eq:p_q_modq}
\end{gather}
and we have
\begin{gather*}
p \left[\dfrac{1}{p}\right]_{q}
+ q \left[\dfrac{1}{q}\right]_{p} =  1 + pq.
\end{gather*}
By letting
\begin{gather*}
\left[-\dfrac{1}{q}\right]_{p} :=   p - \left[\dfrac{1}{q}\right]_{p},
\end{gather*}
and for arbitrary $n \in \ZZ$, 
$\left\{\dfrac{1}{q}\right\}_{p}$ and $\left\{-\dfrac{1}{p}\right\}_{q}$ 
in (\ref{eq:PNR2}) are realized by
$$
 \left\{\frac{1}{p}\right\}_q =
\left[\dfrac{1}{p}\right]_{q} + nq, \quad
 \left\{-\frac{1}{q}\right\}_p =
\left[-\dfrac{1}{q}\right]_{p} + np. \quad
$$
In other words, the number of pairs 
$\displaystyle{\left(\left\{\frac{1}{p}\right\}_q, 
                     \left\{\frac{1}{q}\right\}_p\right)}$
satisfying (\ref{eq:PNR2}) is countably infinite.
As the $n=0$ case is essential, one can identify
them with 
$ \left[\dfrac{1}{p}\right]_{q}$ and $\left[-\dfrac{1}{q}\right]_{p}$
if needs.

\subsection{Quadratic theory}

The  Legendre symbol $ \begin{pmatrix} p \\ s \end{pmatrix}$
for a prime number $s$ is defined by \cite[Chap.5]{IR},
\begin{gather}
\begin{pmatrix} p \\ s \end{pmatrix} :=
\left\{ \begin{matrix} +1, & \text{
if there is an integer $m$ such that $m^2 = p$ mod $s$,}\\
-1, & \text{otherwise.} \end{matrix} \right.
\end{gather}
Further the Jacobi symbol for coprime numbers
$p$ and $q$ is given by
$$
\begin{pmatrix} p \\ q\end{pmatrix}  := 
\prod_{s_1, \cdots, s_n} \begin{pmatrix} p \\ s_i\end{pmatrix} 
$$ 
where  $s_1$, $\cdots$, $s_n$ are prime numbers such that
$q = s_1 \cdots s_n$.

Then the Jacobi quadratic reciprocity is well-known as
\cite[Prop. 5.2.2]{IR},
$$
\begin{pmatrix} p \\ q \end{pmatrix} 
\begin{pmatrix} q \\ p \end{pmatrix} 
=(-1)^{(p-1)/2) (q-1)/2}.
$$

\subsection{Gauss sum}

Here I mention the Gauss sum primitively,
though in the previous article \cite{MO}, it is mentioned in detail.
Though the Gauss sum can be defined using a multiple character and an additive
character generally \cite[Chap. 8 \S 2]{IR}, 
we deal only with the quadratic Gauss sum, which is given by
 \cite[Chap. 6]{IR},
$$
    G(p, q, d):= \sum_{c = 0}^{q-1}
                 \ee^{\frac{2\pi\ii p}{q} (c + d)^2},
$$
where $q$ and $p$ are coprime integers and $d$ is an integer.
(Using (\ref{eq:PNR2}), we sometimes use another expression
$G'(p', q', d):= \sum_{c = 0}^{q'-1} \ee^{\frac{\pi\ii p'}{q'} (c + d)^2}$
as mentioned in \cite{MO} but in this section, we employ the version
with the $2\pi$ prefactor in the exponent\footnote{
The correspondence between both expressions 
is simple for a case $p=2p'$.
For other cases, we need  subtle treatments. When $q'$ is odd case,
we find $\displaystyle{\left[\frac{1}{2}\right]_{q'} \in \ZZ/{q'}\ZZ}$
satisfying (\ref{eq:p_q_modq}). More precise argument is left to \cite{MO}.}.
)
For simplicity, we consider $G(p, q) := G(p, q, 0)$ and the case
that $q$ is an odd prime number.
Since 
$$
     \sum_{c = 0}^{q-1} \ee^{\frac{2\pi\ii p}{q} c} = 0,
$$
we have another representation using the  Legendre symbol,
$$
    G(p, q)= \sum_{c = 0}^{q-1}
             \begin{pmatrix} p c \\ q \end{pmatrix}
                 \ee^{\frac{2\pi\ii p}{q} c}
    = \begin{pmatrix} p \\ q \end{pmatrix} G(1, q), \quad
    G(1, q)= \sum_{c = 0}^{q-1}
             \begin{pmatrix} c \\ q \end{pmatrix}
                 \ee^{\frac{2\pi\ii p}{q} c}.
$$
It is not difficult to prove that \cite[Prop. 6.3.2]{IR},
$$
              G(1, q)^2 = (-1)^{(q-1)/2} q.
$$
Thus it is a concerned problem to determine the sign in the Gauss sum
\cite[Chap.6 \S 4]{IR} and then we obtain
$$
              G(1, q) = \left\{ 
   \begin{matrix} 
    \sqrt{q}, &\mbox{ if } q \equiv 1 \quad \mbox{module }4, \\
   \ii\sqrt{q}, &\mbox{ if } q \equiv 3 \quad \mbox{module }4. \\
   \end{matrix}\right. 
$$
More general case $G(p, q, d)$ is mentioned well in \cite[Appendix]{MO}.

The Gauss reciprocity is studied well due to Hecke \cite[Chap.8]{He},
which is explained in detail in \cite{MO},
\begin{equation}
\frac{1}{|q|^{1/2}} \sum_{c \in \ZZ/q \ZZ}
 \ee^{\frac{\pi\ii p}{q} (c + d)^2}
=
 \ee^{\frac{\pi\ii}{4}\sign(pq)}
\frac{1}{|p|^{1/2}} \sum_{c \in \ZZ/p \ZZ}
 \ee^{-\frac{\pi\ii q}{p} c^2 -2\pi\ii d c} .
 \label{eq:Grecp}
\end{equation}
Due to Weil representation \cite{Wei1}, we could regard that
the factor $\ee^{\pi \ii/4}$ is related to the Maslov index 
and phase anomaly \cite{Sch,BW}

\section{Gauss Optics and $\SL(2, \RR)$}
\label{sec:1}

Here let us review Gauss optics following the Guillemin and Sternberg 
\cite{GS}.

In \cite{G2}, Gauss showed us that the optical system is
recognized as a $\SL(2, \RR) = \Sp(2, \RR)$ map between
incoming plane $S_1$ and outgoing screen $S_2$.
For $c = 1, 2$,
we choose the coordinate systems denoted by
$$
  w_c := \begin{pmatrix} x_c \\ u_c \end{pmatrix} \in S_c,
$$
where $u_c = d x_c / d z$ is the angle variable at 
$S_c$ respectively along the optical axis $z$.
The origin of $x_c$ coincides with the optical axis.
In the Gauss optics, {\it{i.e.}}, two-dimensional linear
optics\footnote{In \cite{G2}, Gauss dealt with three dimensional
optical system $(x, y, z)$ with cylindrical symmetry.
$(x, dx/dz, y, dy/dz, z)$ was dealt with but the cylindrical symmetry
reduces it to two dimensional linear optical system $(r, z)$ or $(r, dr/dz, z)$ 
for $r = \sqrt{x^2 + y^2}$.},
the optical system is represented by
the special linear group 
$$
\fg \in \SL(2, \RR)
  :=\left\{ \begin{pmatrix} A & B \\ C& D \end{pmatrix} \ | \
       A D - B C = 1, \ A, B, C, D \in \RR \right\}
$$
such that its action to $S_c$ is given by,
$$
	w_2 = \fg w_1.
$$
In other words, every 
two-dimensional linear optical system (or Gaussian optical system)
corresponds to an element $\SL(2, \RR)$ bijectively;
$\begin{pmatrix} 1 & \delta z \\ 0& 1 \end{pmatrix}$ corresponds to
the translation by $\delta z$ along the optical axis whereas
$\begin{pmatrix} 1 & 0 \\ P & 1 \end{pmatrix}$ to
the thin lens with power $P$. The optical system consists of
combination of various translations and thin lenses while
$\SL(2, \RR)$ is generated by both matrices for appropriate
$\delta z$'s and $P$'s.
Every element $\fg \in \SL(2, \RR)$ preserves
the symplectic product of 
$$
		\langle{w_1, w_1'}\rangle = x_1 u_1' - x_1' u_1,
$$
{\it{i.e.}}, $\langle{w_1, w_1'}\rangle = \langle{\fg w_1, \fg w_1'}\rangle$.

Now we fix an optical system and thus an element
$\fg \in \SL(2, \RR)$ like (\ref{eq:SL2sp}).
For the the system, we deal with the Lagrangian submanifold \cite[p.34]{GS},
$$
S^{\intf}:=\{(w, \fg w) \ | \ w \in S_1\} \subset S_1 \times S_2.
$$
Since $S^\intf$ is known as two-dimensional manifold \cite{GS},
in the interference phenomena,  we pick up the independent
variables $x_1$ and $x_2$  to express $S^{\intf}$.
In other words, $u_c$ $(c = 1, 2)$ is a function of $x_1$ and $x_2$ as
$$
	u_1 = \frac{x_2 - A x_1}{B}, \quad
	u_2 = \frac{D x_2 - x_1}{B}.
$$
Then the optical length is given by
\begin{equation}
\begin{split}
	L &= \frac{1}{2} \langle{w_1, w_2}\rangle +  z_2 - z_1 \\
	  &= \frac{1}{2B} (D x_2^2 - 2 x_1 x_2 + A x_1^2) + z_2 - z_1. \\
\end{split}
\end{equation}
As in \cite{GS}, we have the wave functions $\psi_c$ over
$S_c$ $(c=1,2)$ under the scalar approximation.
For given $\psi_1$ over $S_1$, we have the image of $\psi_2$
at $S_2$;
\begin{gather*}
\begin{split}
	\psi_2(x_2) &=
        \left( \frac{  \ii }{\lambda B} \right)^{1/2}
        \exp\left( \frac{2 \pi \ii }{\lambda } (z_1 - z_2) \right) \cdot\\
        &\int d x_1 \psi_1(x_1)
       \exp\left(
	  \frac{\pi}{B\lambda} \ii(D x_2^2 - 2 x_1 x_2 + A x_1^2)\right),
\end{split}
\end{gather*}
where $\lambda$ is the wave length.
We introduce $\phi_2$ by the relation,
$$
\psi_2(x_2) = \ee^{\left( \frac{2 \pi \ii }{\lambda } (z_1 - z_2) \right) }
	\phi_2(x_2) .
$$

\section{Talbot phenomena}

In this section,
 we will review the fractional Talbot phenomena \cite{BK,KK,MO,WW1}
and consider a relation between the Gauss sum and the Gauss optics
explicitly.

As in \cite{BK,MO},
 we will consider the $\delta$-comb grating plane $z=0$, 
\begin{gather}
	\psi_1(x) = \sum_{n \in \ZZ} \delta(x - n a).
\label{eq:deltacomb}
\end{gather}
Here we note that there is a group action $\ft_a$ on $S_c$:
\begin{gather}
      \ft_a \cdot \begin{pmatrix} x\\ u\end{pmatrix}
 = \begin{pmatrix} x + a \\ u - \frac{A}{2B} a\end{pmatrix}.
\label{eq:ta}
\end{gather}
The $\delta$-comb gives the distribution at the screen,
\begin{gather}
\begin{split}
\phi_2(x_2) = 
        \left( \frac{  \ii }{\lambda B} \right)^{1/2}
\sum_{n\in \ZZ} \exp\left(
	  \frac{\pi}{B\lambda} \ii(D x_2^2 - 2 n a x_2 + A n^2 a^2)\right).
\end{split}
\label{eq:phi2I}
\end{gather}
We write this by $\phi_2^I$.

Using the Poisson sum relation of (\ref{eq:deltacomb}),
\begin{gather}
	\psi_{1}(x_1,0)=\sum_{n \in \ZZ} 
            \frac{1}{a}\exp\left(2\pi \ii\frac{x_1 n}{a} \right)
	            =\sum_{n \in \ZZ} \delta(x_1 - a n ),
\end{gather}
 we have another expression of (\ref{eq:phi2I}) \cite{KK}
\begin{gather}
\begin{split}
\phi_2(x_2) = 
        \left( \frac{  1 }{A a^2} \right)^{1/2}
 \exp\left( \frac{\pi\ii x_2^2}{ \lambda} C\right) 
\sum_{n\in \ZZ}
 \exp\left(
	  \pi \ii\left(\frac{ 2 n  x_2}{a A} 
         - \frac{B\lambda n^2 }{A a^2} \right)  \right).
\end{split}
\label{eq:phi2II}
\end{gather}
We write this by $\phi^{II}_2$.
In \cite{MO}, we have obtained the essentially same as the expression 
(\ref{eq:phi2II}) using the Helmholtz equation.
As (\ref{eq:phi2II})  comes from the $\frac{1}{a}\exp(2\pi i\frac{x n}{a} )$
which exhibits wavy properties,
we will regard (\ref{eq:phi2II}) as the wavy expression.
This is contrast to (\ref{eq:phi2I}), which we are to consider
as a particle-like expression.

Noting that $a^2/\lambda$ is the order of length and the unit of
the system, we will scale the variables as,
$$
	\hx_c := \frac{x_c}{a}, \quad
	\hu := u, \quad
   \begin{pmatrix} \hA & \hB \\ \hC& \hD \end{pmatrix}
  := \begin{pmatrix} A & B a^2/\lambda \\ C\lambda/a^2& D \end{pmatrix},
\quad \hphi_2^{I, II}(\hx_2) := a \phi_2^{I, II}(x_2).
$$

Then we have the relations,
\begin{gather}
\begin{split}
\hphi_2^I(\hx_2) &= 
        \left( \frac{  \ii }{\hB} \right)^{1/2}
\sum_{n\in \ZZ} \exp\left(
	  \frac{\pi}{\hB} \ii(\hD \hx_2^2 - 2 n \hx_2 + \hA n^2)\right), \\
\hphi_2^{II}(x_2) &= 
        \left( \frac{  1 }{\hA} \right)^{1/2}
 \exp\left( \pi\ii \hx_2^2 \hC\right) 
\sum_{n\in \ZZ}
 \exp\left(
	  \pi \ii\left(\frac{ 2 n \hx_2}{ \hA} 
         - \frac{\hB n^2 }{\hA} \right)  \right). \\
\end{split}
\label{eq:phi2I,II}
\end{gather}
As we mentioned in Introduction, they are written by the elliptic
theta functions \cite[p.35]{P},\cite[p.463]{WW2}. By letting
$$
        \tau := \frac{\hB}{\hA},
$$
$\phi$'s are written by
\begin{gather}
\begin{split}
\hphi_2^I(\hx_2) &= 
        \left( \frac{  \ii }{\hB} \right)^{1/2}
	\ee^{\pi\frac{\hD}{\hB} \ii\hx_2^2} 
          \theta\left(-\frac{\hx_2}{\hB}; \tau\right),\\
\hphi_2^{II}(\hx_2) &= 
        \left( \frac{  1 }{\hA} \right)^{1/2}
 \ee^{\pi\ii \hC\hx_2^2}
	\theta\left(\tau\frac{\hx_2}{\hB}; -\frac{1}{\tau}\right).\\
\end{split}
\label{eq:phi2I,II,theta}
\end{gather}
Here $\theta$ is the well-known theta function \cite[p.35]{P}\cite[p.463]{WW2},
$$
   \theta(u, \tau) := \sum_{n\in \ZZ} \ee^{\pi\ii(2 n u + \tau n^2)}.
$$
The system of the Talbot phenomena 
is written by the elliptic theta function\footnote{
As mentioned in Introduction, this historical meaning is very interesting.}.
Then the equality,
$$
\hphi^{I}_2 = \hphi^{II}_2,
$$
is interpreted as the Jacobi imaginary transformation in the
elliptic theta functions \cite[p.475]{WW2}.

\bigskip

Let us consider the fractional Talbot phenomena in the Gauss
optics and its connection to the Gauss sums and the Gauss
reciprocity. The ordinary Talbot phenomena
was studied in \cite{KK}.

Let us consider the case
\begin{gather}
	\frac{A}{B} \frac{a^2}{\lambda} = 
	\frac{\hA}{\hB}  = \frac{p}{q}, \quad
	\kappa_1:=\frac{\hD}{\hA} , \quad
	\kappa_2:=\frac{1}{\hA}  , \quad
	\kappa_3:=\hC , \quad
   \label{eq:ABcond}
\end{gather}
where $p$ and $q$ are coprime numbers and
$$
\kappa_3 = \frac{p}{q} \left(\frac{\kappa_1}{\kappa_2^2} - \kappa_2\right).
$$
Then we have
\begin{gather}
\begin{split}
\hphi_2^I(\hx_2) &= 
        \left( \frac{  \ii }{\hB} \right)^{1/2}
\sum_{n\in \ZZ} \exp\left(
	  \frac{p}{q} \pi\ii
\left(\kappa_1 \left(\hx_2\right)^2 
  - 2 \kappa_2 n  \hx_2 + n^2 \right)\right),\\
\hphi_2^{II}(\hx_2)& = 
        \left( \frac{  1 }{\hA} \right)^{1/2}
 \exp\left( \pi\ii \kappa_3 \left(\hx_2\right)^2\right) 
\sum_{n\in \ZZ}
 \exp\left(
	  \frac{1}{p}\pi \ii\left( 2 n \kappa_2 p \hx_2 
         - q n^2 \right)  \right).
\end{split}
\end{gather}

For $n = r \ell +  s$, we have
$$
	\frac{1}{r}(K_1 n + t n^2)
	= K_1 \ell + 2 t s \ell +  t r \ell^2 +
\frac{1}{r}(K_1 s + t s^2).
$$ 
and thus
$$
	\sum_{n\in\ZZ} \ee^{\frac{\pi\ii}{r} (K_1 n + t n^2)}
	=\sum_{\ell\in\ZZ} \sum_{s=0}^{r-1}
 \ee^{\pi\ii (K_1  + t r) \ell}
 \ee^{\frac{\pi\ii}{r} (K_1 s + t s^2)}.
$$
Here we used the fact $\ee^{\pi\ii s \ell^2} =\ee^{\pi\ii s \ell}$
and $\ee^{2t s \ell \pi\ii}=1$
for  integers $s$, $t$ and $\ell$.
Using these properties, $\phi_2$'s  become
\begin{gather}
\begin{split}
\hphi_2^I(\hx_2) &= 
        \left( \frac{  \ii }{\hB} \right)^{1/2}
 \ee^{  \ii\pi\frac{p}{q}\left(\kappa_1 \hx_2^2\right)}
\sum_{\ell\in\ZZ} 
\ee^{\pi\ii (2\kappa_2 p \hx_2  + p q) \ell}
	 \sum_{s=0}^{p-1}
 \ee^{\frac{\pi\ii}{q} (2\kappa_2 \hx_2 s  + p s^2)},\\
\hphi_2^{II}(\hx_2)& = 
        \left( \frac{  1 }{\hA} \right)^{1/2}
 \ee^{ \pi\ii \kappa_3 \hx_2^2} 
\sum_{\ell\in\ZZ} 
\ee^{\pi\ii (2\kappa_2 p \hx_2  + p q) \ell}
	 \sum_{s=0}^{q-1}
 \ee^{\frac{\pi\ii}{p} (2 \kappa_2 p\hx_2 s  + q s^2)}.\\
\end{split}
\end{gather}
As in \cite{BK,MO}, the wave function of the
system is rewritten as
\begin{gather}
	\hphi_2^{I, II}(\hx_2)= 
           \sum_{n= -\infty}^\infty
          \cA^{I, II}(n; q, p ) \delta (\kappa_2\hx_2 
                 -\frac{1}{2}e_{q p }
           - \frac{n}{q}  ),
             \label{3-8}
\end{gather}
where
\begin{gather}
	e_{q p} := \left\{\begin{matrix} 1, &\text{if } q p & \text{odd,} \\
	        0, &\text{if } q p & \text{even.} \end{matrix} \right.
           \label{2-13}
\end{gather}
By choosing an appropriate prefactor,  we have
\begin{gather}
\begin{split}
          \cA^I(n; q, p )&=\sqrt{\frac{\ii}{p}}
              \sum_{s = 0}^{p-1}
   \exp\left({\ii\pi  \left[\left(2 n +
           q e_{qp}\right) s  + q s^2\right]/p +
          \hkappa_1
\left(2 n +q e_{q p}\right)^2/4p q }\right), \\
         \cA^{II}(n; q, p )&=\sqrt{\frac{1}{q}}
              \sum_{s = 0}^{q-1}
                 \exp\left({i\pi
         \left[\left( 2n+q e_{qp}\right)
             s - p s^2\right]/q }
          +\hkappa_3
\left(2 n +q e_{q p}\right)^2/4q^2\right),
\label{eq:AI,II}
\end{split}
\end{gather}
where 
\begin{gather}
\hkappa_1:= \frac{\kappa_1}{\kappa_2^2}=\hA \hD, \quad
\hkappa_3:= \frac{\kappa_3}{\kappa_2^2}=\hA^2 \hC.
\label{eq:hkappa}
\end{gather}
Provided that $\hkappa_1$ and $\hkappa_3$ are some
integers (or, more precisely speaking, certain fractional numbers), 
these are merely the Gauss sums.
It implies that there appears the fractional Talbot phenomena
in the Gauss optics, even though \cite{KK} argued only the
integral case or $q/p=1$ case.

We should note that the equality between 
$\hphi^{I}_2$ and $\hphi^{II}_2$ in (\ref{eq:phi2I,II}) means the reciprocity,
\begin{gather}
\hphi^{I}_2 = \hphi^{II}_2, \quad
\cA^{I}_2 = \cA^{II}_2. \quad
\label{eq:phi=phi}
\end{gather}
As shown in \cite{MO}, it means the Gauss reciprocity (\ref{eq:Grecp})
\cite{He}.
In other words, in the case, the Gauss optics, the Gauss sums,
and the Gauss reciprocity are connected in the fractional
Talbot system.

For the ordinary fractional Talbot phenomena case,
\begin{gather}
   \begin{pmatrix} \hA & \hB \\ \hC& \hD \end{pmatrix}
 = \begin{pmatrix} 1 & q/p \\ 0& 1 \end{pmatrix}.
 \label{eq:SL2sp}
\end{gather}
$ \cA^I(n; q, p )$ is given by \cite{MO},
\begin{gather}
\left\{ \begin{matrix}
\begin{pmatrix} p\\ q\end{pmatrix}
\exp\left(\ii
\pi \left[ \dfrac{1}{4}(q - 1)
        -\left(\dfrac{q}{p}
             \left(\left[\dfrac{1}{q}\right]_{p}\right)^2
               - \dfrac{1}{q p}\right )n^2
  \right]\right), \\
\begin{pmatrix} q\\ p\end{pmatrix}\exp\left( -\ii
\pi \left[\dfrac{1}{4}p
+\left(\dfrac{q}{p}
             \left(\left[\dfrac{1}{q}\right]_{p}\right)^2
               - \dfrac{1}{q p}\right )n^2
               \right] \right),\\
\begin{pmatrix} q\\ p\end{pmatrix} \exp
      \left(-\ii \pi
          \left[ \dfrac{1}{4}p\qquad\qquad\qquad
             \qquad\qquad\qquad \right.\right.& \\
         \qquad\qquad \left.\left.
           +\left(\dfrac{2q}{p}
               \left[\dfrac{1}{2} \right]_{p}
                \left[\dfrac{1}{2q}\right]_{p}
             -\dfrac{1}{4q p}\right )(2n+q)^2
          \right]
          \right),\\
\end{matrix}
\right. 
\label{eq:AI}
\end{gather}
whereas $ \cA^{II}(n; q, p )$ is given by
\begin{gather}
\left\{ \begin{matrix}
\begin{pmatrix} p\\ q\end{pmatrix}
\exp\left(\ii
\pi \left[ \dfrac{1}{4}(q-1)
         + \dfrac{p}{q}
             \left(\left[\dfrac{1}{p}\right]_{q}\right)^2n^2
  \right]\right),\\
\begin{pmatrix} q\\ p\end{pmatrix}\exp\left(- \ii
\pi \left[\dfrac{1}{4}p-
 \dfrac{p}{q}
  \left(\left[\dfrac{1}{p}\right]_{q}\right)^2 n^2
               \right] \right),\\
\begin{pmatrix} p\\ q\end{pmatrix} \exp\left(\ii
\pi \left[ \dfrac{1}{4}(q - 1)
         + \dfrac{2p}{q}
\left[\dfrac{1}{2}\right]_{q}
  \left(\left[\dfrac{1}{2p}\right]_{q}\right)^2
      (2n+q)^2
             \right]\right),\\
\end{matrix}
\right.
\label{eq:AII}
\end{gather}
where both are for {\lq\lq}$p$ even, $q$ odd{\rq\rq}, 
{\lq\lq}$p$ odd, $q$ even{\rq\rq}, and
             {\lq\lq}$p$ odd, $q$ odd{\rq\rq} 
respectively.

\bigskip

\section{Heisenberg Group and Schr\"odinger representation}

Here we review the Weil representation in order to
answer the question why the Gauss sum appears in the optical system.

Let us consider a ring $R$ and $S := {}^t(R, R)$.
We assume that $R$ is $\RR$, $\ZZ$, $\QQ$ or $\ZZ/b \ZZ$,
where $b$ is a positive odd number.
The case that $R = \RR$ is studied well in \cite{GS,RS}  
for the optical system based upon \cite{LV,Wei1}
and thus so in this article, we basically assume 
 that $R$ is $\ZZ/b \ZZ$.
When $R$ is the finite ring, 
the Weil representation and Heisenberg group are recently
studied well \cite{Bl,CMS,Sc,Sz}.
In this article, we will consider only the simplest case and
so if one considers more complicate cases, 
 \cite{Bl,CMS,Sc,Sz} are nice for the purpose and may
provide the guide.

\subsection{Heisenberg Group}

Let us consider the Heisenberg group $H$ associated with
 $S=R^2$ and  $Z= R$ \cite{Bl,CMS,Sc,Sz},
$$
      H :=  (S, Z) 
$$
with the product $H \times H \to H$ defined by
$$
	(\hw_1, z_1) (\hw_2, z_2) =
	(\hw_1 + \hw_2, z_1+z_2+\frac{1}{2}
      \langle{\hw_1, \hw_2}\rangle ),
$$
for $((\hw_1, z_1),(\hw_2, z_2)) \in H \times H$.
It is obvious that  the product is well-defined and it becomes a group.

The Heisenberg group is also characterized by an central
extension of the Abelian group (free $R$-module) as \cite[p.17]{P},
$$
	0 \to Z \to  H	\to R^2 \to 0,
$$
such that  $e: R^2\times R^2 \to Z$  by symplectic product
$ \frac{1}{2} \langle{\hw_1, \hw_2}\rangle$
for $((\hw_1, z_1),(\hw_2, z_2)) \in H \times H$.

Then we have the following facts:
 
\begin{enumerate}
\item For $h_1, h_2 \in H$,
$h_2 h_1 h_2^{-1} = (\hw_1, z_1 + \langle{\hw_1, \hw_2}\rangle ).$

\item $N:=\{((0, u), z) \ | \ x, z \in R\}$ 
is a normal Abelian subgroup of $H$.

\item $Z:=\{((0, 0), z) \ | \ z \in R\}$,
 $U:=\{((0, u), 0) \ | \ u \in R\}$,
and $X:=\{((x, 0), 0) \ | \ x \in R\}$ 
are normal Abelian subgroups of $H$ respectively.

\item $H = N \rtimes X$.
\end{enumerate}

For $\gamma \in R^\times$, we have the action $R^\times$ on $H$, 
$$
	\alpha_\gamma\cdot(w, z) = (\gamma w, \gamma^2 z),
$$
which is regarded as an element of $\Aut(H)$.
On the other hand,  $\fg \in \SL(2, R)$ induces the automorphism
of $H$,
$$
	\fg\cdot(w, z) = (\fg w, z),
$$
or $\SL(2, R) \subset \Aut(H)$. 
When $R = \RR$, we have $\SL(2, R) \cap R^\times = \{ \pm 1\}$.
For $R = \RR$ case, we have an exact sequence of topological groups,
$$
	1 \to R^\times \to \Mp(2, R) \to \Sp(2, R) \to 1,
$$
where $\Mp(2, R)$ is the metaplectic group.

When $R$ is a finite ring $\ZZ/b \ZZ$, 
were $b$ are a positive integer,
the Heisenberg group becomes a finite group.
We will consider the automorphism in the group ring $\CC[H]$.

\subsection{Character of Heisenberg Group}

When we regard $\CC[H]$ as $\CC[N]$-module,
we apply the Mackey theory of the finite group \cite{CR1,CR2,Se} to it.
We recall the Mackey theory which is given as follows:

\begin{proposition}
Let $K$ is an arbitrary field and $G$ be a finite group.
Let $M$ be a simple $K[G]-module$
and $H$ be a normal subgroup of $G$.
As $M$ can be regarded as $K[H]$-module, we denote it by
$M_H$. 
Then followings hold

\begin{enumerate}

\item
$M_H$ is  completely reducible.

\item
The irreducible $K[H]$-submodules of $M_H$
are all conjugates of each other.
$$
M_H \approx L^{(g_1)} \oplus
 L^{(g_2)} \oplus \cdots \oplus L^{(g_r)}.
$$

\item There are a subgroup $S$ of $G$, called
\textrm{inertia group}, and $K[H]$-module $L$ such that
for $ g_i \in S$, $g_i L = L^{(g_i)}$ and $|S|=r$.
\end{enumerate}
\end{proposition}

We have its character 
$\varrho_\eta: Z \to \CC^\times$ parameterized by $\eta \in \RR$,
{\it{ e.g.}}, $\eta = b$,
\begin{equation}
	\varrho_\eta(z) =\exp\left( \frac{2\pi}{\eta}\ii z \right).
\label{eq:varrho}
\end{equation}
As $N$ is Abelian,
the natural projection $\varpi : N \to Z$ is
 a group homomorphism and thus we define 
$\varrho_\eta: N \to \CC^\times$ by,
$$
	\varrho_\eta(n) := \varrho_\eta \circ \varpi (n) .
$$
For $n \in N$ and $h' \in H$, we have a natural action
on $\varrho_\eta \in \Hom(Z, \CC^\times)$,
$$
(h \circ \varrho_\eta)(n) :=
\varrho_\eta( h' \cdot n \cdot {h'}^{-1})
=\exp\left(
 \frac{2\pi}{\eta}\ii \left( z + \langle{w', w}\rangle\right) \right).
$$
Noting $\langle{X, U}\rangle \not \equiv 0$, and
 $\langle{X, X}\rangle = \langle{U, U}\rangle = 0$,
we may regard  that $X$ has an action on 
$N{\ \hat{}}:= \Hom(N, \CC^\times)$.

For $\varrho_\eta \in N{\ \hat{}}/X$, we consider $X_\varrho (\subset X)$ as
the stablizar to $\varrho_\eta$.
When $\varrho_\eta$ is trivial,  $X_\varrho$ is equal to
$X$ and then, the representation becomes $R^2$.

On the other hand, if $\varrho_\eta$ is a non-trivial case,  
$X_\varrho$ is equal to $\{0\}$ and then
we consider the induced representation $ \ind_N^H(\varrho_\eta)$.

We should note that $N$ is a normal subgroup 
$N \triangleleft H$ and thus we apply the Proposition
to this system,
$$
\CC[H] \approx \CC[N_1] \oplus \CC[N_2] \oplus 
\cdots \oplus \CC[N_{b^2-1}] \oplus \CC[N_{b^2}]
= \oplus_{x \in X} \CC[x N].
$$
Here $\CC[N_i]$ is $\CC[N]$-module and $X$ is the inertia group.
For $ h = x n $ of $x\in X$, $n \in N$, we have
$$
\varrho_i(h) = \tr \cS(h) := \left\{
         \begin{matrix} \varrho_\rho(z) & \mbox{	for } h=(0,0, z) \\
         0 & \mbox{	otherwise } \end{matrix} \right. . 
$$
We will consider a function over  $H$, or
an element of $(\chi(0), \chi(1), \cdots, \chi(b-1))$
belonging to $\oplus_{x \in X} \CC[x N] \approx \CC[H]$.
By checking the action of $X$, $U$ and $Z$,
we have the Schr\"odinger representation of $H$ which is generated by
$$
	\cS(x) \chi((x',u',z')) = \chi((x + x', u', z')) ,
$$
$$
	\cS(u) \chi((x',u',z')) = \ee^{\frac{2\pi\ii}{\eta} u x'}
                \chi((x', u', z')) ,
$$
$$
	\cS(z) \chi((x',u',z')) = \ee^{\frac{2\pi\ii}{\eta} z }
                \chi((x', u', z')) ,
$$
or
\begin{equation*}
    \cS(x) = \begin{pmatrix} 
    0 & 1 &   &       &   &  \\
      & 0 & 1 &       &   &  \\
      &   & 0 & \ddots&   &  \\
      &   &   & \ddots& 1 &  \\
      &   &   &       & 0  & 1 \\
    1 &   &   &       &  &   0 \end{pmatrix}^x, \quad
    \cS(u) = \begin{pmatrix} 
 1 &                  &  &     \\
   &\ee^{2\pi\ii u/\eta}&  &     \\
   &                  &\ddots&   \\
   &                  &      & \ee^{2\pi(b-1) \ii u/\eta}
 \end{pmatrix}.
\end{equation*}

When $R=\RR$, we consider $L^2(\RR)$ instead of $\CC[X]$ and then
we could define $dW(\xi)$, $\xi \in \mathfrak h$
  for the Lie algebra $\mathfrak h$ of $H$,
$$
	d\cS(\xi) \chi= \frac{d}{d t} \cS(t \xi) \chi|_{t=0}.
$$
Due to \cite[15]{LV}, we have
\begin{equation}
	d \cS(\xi_x) = \frac{d}{d x}, \quad
	d \cS(\xi_u) = \frac{2 \ii \pi}{\eta} x, \quad
	d \cS(\xi_z) = \frac{2 \ii \pi}{\eta} id. \quad
   \label{eq:CCRW0}
\end{equation}
This means that
 $\cS(x)= \ee^{x d\cS(\xi_x)}$, $\cS(u)= \ee^{u d\cS(\xi_u)}$, and
 $\cS(z)= \ee^{zd\cS(\xi_z)}$.
We have the canonical commutation relation,
\begin{equation}
 [ d \cS(\xi_x),  d \cS(\xi_u)] =   d \cS(\xi_z),
\quad \mbox{or}\quad
\frac{d}{dx} x - x \frac{d}{dx}  = 1. 
   \label{eq:CCRW}
\end{equation}

\subsection{Representation of Automorphism of Heisenberg Group}

We mention the representation of $\Aut(H)$ for the Schr\"odinger
representation $\cS$.
First we deal with
the action  $\alpha_\gamma$ of $R^\times$,
$$
   \alpha_\gamma \cdot \cS(h) = \cS( \alpha_\gamma(h))
    = \gamma^{2}\cS(h).
$$
Secondary we consider the $\fg \in \SL(2, R)$.
By letting
$$
\fg \circ \cS(h) = \cS(\fg^{-1} h) ,
$$
it is shown that there exists the unitary action $W(\fg)$
on $H$ such  that
$$
\cS(\fg h) = \cW(\fg) \cS(h)  \cW(\fg)^{-1}, 
$$
for every $h \in H$ when $\CC[H]$ is regarded as $\CC[H]$-module.
By tuning the factor, we obtain
 the Weil-representation of the metaplectic group $\Mp(2, R)$.

Following the case $R=\RR$  \cite[(3.15)]{RS},
the Weil representation $\left[\cW(\fg) \chi\right](\hx_2)$ is given by
\begin{gather}
\begin{split}
& \sum_{\hx_1 \in R} G(\hx_2; \hx_1) \chi(\hx_1) \\
& = 
\sqrt{\frac{A\ii}{B\eta}}
\sum_{\hx_1 \in R, (\hx_2, \hu_2) = \fg(\hx_1, \hu_1)} 
\exp\left(
\frac{2\pi}{\eta}\ii \left( 
\frac{1}{2}\langle{(\hx_1, \hu_1), (\hx_2, \hu_2)}\rangle\right) \right)
\chi(\hx_1),
\label{eq:Weilrep}
\end{split}
\end{gather}
where  $\hu_c = \hu_c(\hx_1, \hx_2)$ $(c=1,2)$.
Here $G(\hx_2; \hx_1)$ has its multiplication 
$$
G(\hx_3; \hx_1)  
 = \sum_{\hx_2 \in R} G(\hx_3; \hx_2) G(\hx_2; \hx_1).
$$
The phase of $\displaystyle{ \frac{\hA \ii}{\hB\eta}}$
(\ref{eq:Weilrep}) is given by 
$$
	s(\fg) = \sign(\hB) \ee^{\pi\ii/2}.
$$
Hence $\cW$ is the representation of
the metaplectic group $\Mp(2, R)$.

We note that for example as in the path integral \cite{Sch},
the computation of the kernel function $G(\hx_2; \hx_1)$ is based upon
the canonical commutation relation (\ref{eq:CCRW}). 

\bigskip

\section{Gauss sum in fractional Talbot phenomena, revised}
\label{sec:revised}

In this section, we will investigate
the relation between the Gauss sum and the fractional Talbot phenomena
again more algebraically.
 This is a revised investigation of \cite{MO}.
In other words,
we consider why the optical system is expressed by the
Gauss sums.
We have to consider the symmetries of
the system which insert the discrete pictures in
the optical system and 
give an answer the question why 
(\ref{eq:CCR}) is similar to (\ref{eq:PNR}).

Let us reconsider the physical situations in \S 2 and \S 3.

\subsection{The translation action $\ft_a^\fg$ }

Here we will consider the first discrete
nature in the Talbot phenomena coming from the delta-comb slit;
due to it, the system is represented  by  theta function.
Let us fix $\fg \in \SL(2, \RR)$, which means that we choose
an optical system.

As the Lagrange submanifold $S^\intf$ is now two-dimension, 
the parameters $\hx_1$ and $\hx_2$ of $\hw_2 = \fg \hw_1$ are 
its local coordinates of $S^\intf$ and
thus $\hu_c$ is expressed by \cite[p.34]{GS},
$$
	\hu_c = \hu_c(\hx_1, \hx_2, \fg), \quad \mbox{for  } c = 1, 2.
$$

We are concerned with the interference system $S^\intf$
with the translation symmetry (\ref{eq:ta}).
The action of translation (\ref{eq:ta}) induces
$$
      \begin{pmatrix}
\hx_1 \\ \hu_1
       \end{pmatrix} = 
	(\ft_a^{\fg})^n
      \begin{pmatrix} \hx_0 \\ \hu_0 \end{pmatrix} = 
      \begin{pmatrix} \hx_0 \\ \hu_0 \end{pmatrix} +
      \begin{pmatrix}  n \\  - \frac{\hA}{\hB}  n \end{pmatrix} ,
$$
so that it preserves $\hx_2$ as $x$ component
of image of $\fg$, {\it{i.e.}},
$$
      \begin{pmatrix} \hx_2 \\ \hu_2 \end{pmatrix} = 
	\fg\cdot (\ft_a^{\fg })^n \begin{pmatrix} \hx_0 \\ \hu_0\end{pmatrix}
	=\fg \begin{pmatrix} \hx_0 \\ \hu_0 \end{pmatrix}
        - \begin{pmatrix} 0 \\ \frac{1}{\hB}   n\end{pmatrix},
$$
which provides
$$
\frac{1}{2} \langle
 \ft_a^{\fg n} \hw_1, \fg \ft_a^{\fg n} \hw_1
\rangle = \frac{1}{2\hB} 
   (D \hx_2^2 - 2 (\hx_0 + n ) \hx_2 + \hA (\hx_0 + n )^2) .
$$

The above translation means that we deal with 
$$
H^{(a, \fg)}_{\hw_0} := \{ (\hw, z) \ | \ w 
= \ft_a^{\fg \ n} \hw_0, \  n  = 0, 1, \cdots, b-1\}.
$$
over $R = \QQ[[\hA, \hD, 1/\hB, x_0, u_0]]$,
a formal power series of $\hA, \hD, 1/\hB, x_0$ and $u_0$ over $\QQ$.
Here we should note that for $\hw = (0, 0)$ case,
$H^{(a, \fg)}_{0} := \{ (n, - \hA/\hB n, z) \ $ $ | \ n \in \ZZ, z \in R\}$
is a normal subgroup of $H$ over $R = \QQ[[\hA, \hD, 1/\hB, x_0, u_0]]$.
$\CC[H^{(a, \fg)}_{\hw_0}]$ is $\CC[H^{(a, \fg)}_{0}]$-module.
Hence this discrete system consists of $H$ itself.
The translation does not break the algebraic structure of
the optical system given by \cite{GS,RS}.

From (\ref{eq:phi2I,II,theta}),  we have the theta function
expression due to the symplectic structure.
In fact, 
$ \begin{pmatrix}  n \\  - \frac{\hA}{\hB}  n \end{pmatrix}$
is written by 
$ \begin{pmatrix}  n \\  - \tau  n \end{pmatrix}$
which shows the periodic structure in the Abelian variety of
genus one \cite{P}\footnote{It is very interesting
that Talbot himself studied the Abelian integral \cite[p.413]{C}.}.
Oskolkov \cite{Osk} and Berry and Bodenschatz \cite{BB}
dealt with different $\tau$ as time development
and showed  interesting patterns.

\subsection{Discrete nature in the optical system}

Here we will consider the second discrete
nature in the Talbot phenomena.
In order to insert another discrete nature $\QQ$ in this system, we have
imposed the condition (\ref{eq:ABcond}),
\begin{gather}
	\frac{\hA}{\hB}  = \frac{p}{q}, \quad
\label{eq:ABcond2}
\end{gather}
where $p$ and $q$ are coprime numbers.
Let us consider realization of 
(\ref{eq:ABcond})  or (\ref{eq:ABcond2})  in $\SL(2, \ZZ)$ and then 
we naturally encounter the simplest case,
$$
\begin{pmatrix} p & q \\
	\left\{-\frac{1}{q}\right\}_p &\left\{\frac{1}{p}\right\}_q
     \end{pmatrix} \in \SL(2, \ZZ),
$$
where
\begin{equation}
\det\begin{pmatrix} p & q \\
	\left\{-\frac{1}{q}\right\}_p &\left\{\frac{1}{p}\right\}_q
     \end{pmatrix}
 = p \left\{\frac{1}{p}\right\}_q
	-q \left\{-\frac{1}{q}\right\}_p = 1.
\label{eq:detSLZ}
\end{equation}
This recovers (\ref{eq:PNR}) and then (\ref{eq:hkappa})  becomes
$$
	\kappa_1 = p \left\{\frac{1}{p}\right\}_q, \quad
	\kappa_3 = p^2 \left\{-\frac{1}{q}\right\}_p, \quad
$$
and then $\mathcal A_I$ is represented by
 the Gauss sum explicitly.

The above condition  corresponds to the ordinary fractional Talbot phenomena
case (\ref{eq:SL2sp}) in $\SL(2, \QQ) \subset \SL(2, \RR)$,
$$
	\begin{pmatrix} 1 & \hat q/p \\ 0 & 1 \end{pmatrix}
=\begin{pmatrix} \frac{1}{p} & 0 \\
	\left\{\frac{1}{q}\right\}_p & p \end{pmatrix}
\begin{pmatrix} p & q \\
	\left\{-\frac{1}{q}\right\}_p &\left\{\frac{1}{p}\right\}_q
     \end{pmatrix} \in \SL(2, \QQ).
$$
Using this fact, we will give an answer to the question in Introduction
and \cite{MO} in next section.

\bigskip
\subsection{The fractional Talbot phenomena and Weil representation}

By inserting the discrete nature with the translation properties
and $\SL(2, \ZZ)$ into the Gauss optics in the previous subsections, 
we encounter $\ee^{2\pi\ii/q}$.

Here we will give its connection with the Weil representation in the
previous section in order to consider the Gauss sum again.

Suppose that $R=\ZZ/q\ZZ$, and $\hx_0$, $\hu_0$  are elements of $R$.
For simplicity, $q$ is an odd number.
The character $\chi_q$ of $R$ is given by $\ee^{2\pi\ii/q}$.
In order to choose $\hB \in R^\times$ freely,
we restrict the group $\fg$ belonging to
$$
\Gamma(2, R) := \left\{\fg:=\begin{pmatrix} \hA & \hB \\ 0 & \hC\end{pmatrix}
        \ | \ \fg \in \SL(2, R)\right\},
$$
and we set $\hA/\eta \hB = q/p$.
More specially, when we set
\begin{gather*}
\fg=\begin{pmatrix}1 & 1\\0 & 1\end{pmatrix}, 
\quad \eta = \frac{q}{p},
\quad \chi(\hx_1) \equiv 1\text{  for every }\hx_1,
\end{gather*}
$\left[\cW(\fg) \chi\right](\hx_2)$ in 
(\ref{eq:Weilrep}) is equal to
\begin{gather}
\sqrt{\frac{q\ii}{p}}
\sum_{\hx_1 \in R}
\exp\left(
 \frac{q\pi}{p}\ii 
   (\hx_2^2 - 2 (\hx_0 + n ) \hx_2 + (\hx_0 + n )^2) \right).
\label{eq:Weilrep2}
\end{gather}
By letting $\hx_2 = (qe_{pq} - 2 n )/ 2q$, 
this is essentially the same as $\mathcal A_2^I$
in (\ref{eq:Weilrep2})  of
$\fg=\begin{pmatrix}1 & q/p\\0 & 1\end{pmatrix}$.

Then we realize $\mathcal A_2^I$ as in (\ref{eq:Weilrep2}).
Using the reciprocity for Gauss sums (\ref{eq:Grecp})
or the reciprocity corresponding to the wave-particle
complementarity (\ref{eq:phi=phi}), 
we also realize $\mathcal A_2^{II}$ (\ref{eq:AII}).

\bigskip

\section{Discussion}

In this article, we dealt with the Gauss optics with
 the delta-comb and we gave explicit expressions in terms
of the theta functions (\ref{eq:phi2I,II,theta}).
After considering the fractional condition,
$$
	\frac{\hA}{\hB} = \frac{p}{q} \in \QQ,
$$
we expressed the fractional Talbot phenomena in the Gauss optics
 on $\fg \in \SL(2, \QQ)$ explicitly  as in 
(\ref{eq:AI,II}), and gave their relations to 
the Gauss sums and the Gauss reciprocity.
Due to the $\SL(2, R)$ treatment which corresponds to
the Gauss optics, we could argue the  Weil representation
and the Heisenberg group in the optical system \cite{GS,RS}.

When $R$ is continuous case or $\RR$, the Heisenberg group
is a Lie group and we have its Lie algebra.
 In the Schr\"odinger representation,
the Lie algebra is generated by $x, \frac{d}{dx}$ and $1$
with the canonical commutation  relation as the generating
relation of the algebra (\ref{eq:CCRW}) \cite{LV},
\begin{equation}
    \frac{d }{d x} x - x \frac{d}{dx} = 1.
\label{eq:CCR3}
\end{equation}
The relation governs the kernel functions and  automorphism 
of the Heisenberg group like $G(\hx_2; \hx_1)$ in (\ref{eq:Weilrep}).
The automorphism corresponds to the dynamics and time development 
in the quantum mechanics; in the case of the optics, it 
corresponds to the translation along the optical axis.
Thus (\ref{eq:CCR3}) is the fundamental relation in the automorphism.

On the other hand, when $R$ is $\ZZ/q \ZZ$ case, 
the Heisenberg group is a finite group and thus we can not deal with
its infinitesimal difference neither its  Lie algebra. 
Thus we must directly consider the automorphism of the Heisenberg group.
Instead of the canonical commutation relation (\ref{eq:CCRW}),
we have the relation (\ref{eq:detSLZ}),
\begin{equation}
  p \left\{\frac{1}{p}\right\}_q
	-q \left\{-\frac{1}{q}\right\}_p = 1,
\label{eq:FNR3}
\end{equation}
as the fundamental relation of the automorphism of the Heisenberg group,
$\SL(2, \ZZ) \subset \Aut(H)$. 
As the effect of wavelength $\lambda$ is normalized in the relation 
(\ref{eq:ABcond}), 
in (\ref{eq:FNR3}) the wavy properties as the interference condition 
are built in.
Thus (\ref{eq:FNR3}) implicitly connects the linear 
optical property, {\it i.e.}, of an element of
$\SL(2, R)$, and the wavy property, {\it i.e.}, $R=\ZZ$.  

Hence we conclude that behind the fractional Talbot phenomena,
these relations (\ref{eq:CCR3}) and (\ref{eq:FNR3})
exist and both play the same role essentially for the continuous
case and for the discrete case.
This means that we find the answer to the question in \cite{MO}.

\bigskip

We expect that this algebraic treatment of the Talbot phenomena 
has some effects on several fields related to quantum mechanics,
 the optical system, and missing relations between quantum mechanics
and arithmetic theory \cite[p.149]{Ma}\cite{VVZ}.

As in the adelic consideration \cite[Introduction]{VVZ}, the $p$-adic 
quantum mechanics and ordinary quantum mechanics are treated equivalently.
Then our interpretation of the relation between (\ref{eq:CCR3}) and 
(\ref{eq:FNR3}) is consistent with the philosophy of the adelic 
consideration because in $p$-adic quantum mechanics, $p$ is the small
parameter associated with the $p$-adic differential operator \cite{VVZ}.

Further as in survey of Polishchuk \cite{P}, the Gauss sum 
and the symplectic structure determine the structure of the 
Abelian variety though the theta functions. 
Due to the properties of the Abelian structure,
{\it i.e.}, theorem of cube, the Gauss sum is connected with
another physical problem, Chern-Simons-Witten theory of
the three-manifold related to some Riemann surfaces\cite{De,J,Tu}.
Recently the Abelian variety (more precisely Jacobi variety),
we have explicit representations \cite{MP}.
Using the recent developments and our new result of the fractional
Talbot phenomena, we could investigate the quantum structure 
over there.

On the other hand, $\theta$ function appears in my recent work on a statistical
mechanical problem of closed elastic curves in a plane \cite[Remark 4.3]{M2},
which is closed related to the integrable system. As in the integrable
system, the symplectic structure plays important roles there \cite{GS}.
As mentioned in \cite[Introduction]{M2} in detail,
the problem might be related to $\SL(2, \ZZ)$ in replicable
function theory and monstrous moonshine phenomena \cite{JM};
it might be also associated with another physical problem, the 
Witten 24-manifold. There the concrete relation among 
the symplectic structure and $\SL(2, \ZZ)$ are also one of the 
central theme of the studies \cite{JM,M2}.
The elastic curve problem could be extended as higher dimensional
objects using the Dirac operator case as in \cite[references therein]{M1}. 
Even in the case, the theta functions are defined using the integrable
system and then we should consider a connection between symplectic
structure and wave properties when we consider some quantization
\cite{M0}.

I believe that my interpretation of the fractional Talbot phenomena must
have crucial effects on these studies.

\begin{acknowledgements}
I thank  K. Tamano, N. Konno and H. Mitsuhashi for
the lectures related to \cite{CR1} at Yokohama national university,
discussions, and continuous encouragements. I am also grateful to Y. \^Onishi
for some discussions and encouragements and J. McKay for 
some discussions and telling me the reference \cite{Ma}.
Further I appreciate the referees for critical suggestions and
references \cite{BCSG,FHO,RTCM,WW1}.
\end{acknowledgements}



\end{document}